\documentclass[11pt]{article}
\usepackage{anysize}
\marginsize{3cm}{3cm}{3cm}{3cm}
\usepackage{amsmath,amssymb,bm,caption,graphicx,color,natbib,multirow,amsthm}
\usepackage{mathtools}

\title{Bayesian Non-Parametric Inference for \\ Infectious Disease Data}

\author{Edward S. Knock and Theodore Kypraios\footnote{{\em Address for correspondence:} School of Mathematical Sciences, University of Nottingham, University Park, Nottingham, NG7 2RD, UK. e-mail: theodore.kypraios@nottingham.ac.uk}, \\University of Nottingham, UK}

\begin{document}
\maketitle

\begin{abstract}
We propose a framework for Bayesian non-parametric estimation of the 
rate at which new infections occur assuming that the epidemic is partially observed. The 
developed methodology relies on modelling the rate at which new 
infections occur as a function which only depends on time. Two different types of
prior distributions are proposed namely using step-functions and B-splines. 
The methodology is illustrated using both simulated and real datasets 
and we show that certain aspects of the epidemic such as seasonality and 
super-spreading events are picked up without having to explicitly incorporate 
them into a parametric model.
\end{abstract}

%\begin{keywords}
%non-parametric inference; epidemic; B-spline; changepoint analysis; Bayesian.
%\end{keywords}

\section{Introduction}
%say something general about epidemics
Understanding the spread of communicable infectious diseases in order to prevent  future major outbreaks is high on the global scientific agenda, including contingency planning for the threat of a possible influenza pandemic. The past two decades have seen significant growth in the field of mathematical modelling of communicable diseases and this has led to a substantial increase in our understanding of infectious disease epidemiology and control.

% say something about inference for epidemics being non-trivial.
The transmissible nature of infectious diseases makes them fundamentally different from non-infectious diseases, and therefore the analysis of disease outbreak data cannot be tackled using conventional statistical methods. This is mainly due to the fact that the data are i) highly dependent and ii) incomplete in the sense that the actual transmission process is not directly observed. Although there have been methods developed for parametric estimation under a Bayesian framework (see the recent review by \cite{On10}), little has been done in the area of non-parametric inference.  That is, drawing inference without making certain parametric modelling assumptions for the quantities which govern transmission: i) the force of infection and ii) the period during which an individual remains infectious.

% give some background information
Consider, for example, arguably the most well-known stochastic model in epidemic theory, the so-called susceptible-infected-removed model. In this model individuals are assumed to be in one of three states at any time $t$ during the outbreak: susceptible, infected or removed. In particular, a susceptible individual, after a successful contact with an infectious individual, becomes infected (and infectious) and then after some period of time develops immunity to the disease by entering the removed class. One of the most common assumptions is that the net rate of new infection at some time $t$ ($\lambda_t$) is assumed to be proportional to the density of susceptible individuals ($X_t$) multiplied by the density of infectious individuals ($Y_t$) at time $t$, giving $\lambda_t = \beta X_tY_t$, known as the {\em mass-action principle}. Furthermore, the rate at which individuals recover is often taken as $\gamma Y_t$.  Despite the attempts that have been made to relax this assumption motivated by certain applications (e.g. sexually transmitted diseases), the vast majority of them have been concerned with the assignment of different (parametric) functional forms to $\lambda(t)$ such as $X_tY_t^{\alpha}$ and $X_tY_t/(1 + \alpha Y_t)$ for some unknown parameter $\alpha$ which needs to be estimated from the data; see for example, \citet{OnWen12} and \citet{CapSer78}.

% benefits on non-paramtric inference
The advantages for considering non-parametric inference are manifold. It avoids the false conclusions and biased results which can arise from fitting a given particular model, which can often be selected for simplicity rather than adequacy. It also offers greater model flexibility than parametric models, often constrained by their small number of parameters. Parametric models tend to assume homogeneity of individuals on some level, for example in their susceptibility, infectivity, or in the manner in which they mix within a well-specified population. On the other hand, with a non-parametric approach we are able to allow not just for heterogeneities between individuals, but also variability in the behaviour of an individual and their involvement in the epidemic over time.

\citet{Becker:1989} considered frequentist non-parametric inference of the infection rate in an epidemic by modelling the individual-to-individual contact rate  as a step function with kernel smoothing via a martingale approach. Application was made to the removal time data from the 1967 smallpox epidemic in Abakaliki, Nigeria, effectively under the assumption that infection times were also known by assuming known fixed length latent and infectious periods. Meanwhile, \citet{Kenah:2013} considered a Susceptible-Exposed-Infected-Removed epidemic model by taking a non-parametric survival analysis approach to contact intervals, i.e. the time between an individual becoming infectious and their contacting a given individual. It was assumed that infection times, infectiousness onset times and removal times are known. \citet{Boys:2007} considered a semi-parametric approach by allowing for a time-inhomogenous removal rate modelled using a time--dependent step function. The authors adopted a Bayesian approach and they found that choice of time-homogeneous or -inhomogeneous removal rates affects not only estimation of the removal process, but estimation of the contact process as well.

In this paper we focus on the force of infection and we propose a novel framework which enables non-parametric estimation of the rate at which new infections occur ($\lambda_t$) assuming that only removal data are available. The key idea to our methodology is to treat $\lambda_t$ as a function which only depends on time $t$ and infer such a function within a Bayesian framework. That involves assigning a flexible prior (model) for $\lambda_t$ and we consider two different approaches to model $\lambda_t$, namely a step-function and a 2nd order B$-$spline wherein parameters are estimated using efficient data-augmentation Markov Chain Monte Carlo algorithms.

\section{The Susceptible--Infective--Removed Epidemic Model}
\subsection{Notation and Model Formulation}
Recall the General Stochastic Epidemic model. Consider a population of $N$ individuals, each of whom, at any given time $t\in\mathbb{R}$, is in one of three states: susceptible, infective or removed. The epidemic is initiated by one infective in an otherwise entirely susceptible population. For $t \geq 0$ and $i = 1, \ldots, k$ denote by $X_t$ and $Y_t$ the numbers of susceptibles and infectives at time $t$ respectively. The epidemic process $\{X_t, Y_t\}$ can be defined as a bivariate Markov chain with the following transition rates:
\begin{eqnarray*}
(i,j) \rightarrow (i-1, j+1) & : & \beta X_t Y_t\\
(i,j) \rightarrow (i, j-1) & : & \gamma Y_t
\end{eqnarray*}

and the corresponding transition probabilities to an infection and removal:
\begin{eqnarray*}
\mathbb{P}[X_{t+\delta t} - X_t = -1, Y_{t + \delta t} - Y_t = 1 \mid \mathcal{H}_t] & = & \beta X_t Y_t + o (\delta t)\\
\mathbb{P}[X_{t+\delta t} - X_t = 0, Y_{t + \delta t} - Y_t = -1 \mid \mathcal{H}_t] & = & \gamma Y_t + o (\delta t).
\end{eqnarray*}
All other transitions having probability $o(\delta t)$ and $\mathcal{H}_t$ is the sigma-algebra generated by the history of the process up to time $t$.

It follows that while there is at least one susceptible and at least one infective, new infections in the population as a whole occur at the points of a non-homogeneous Poisson process with rate $\beta X_t Y_t$ and infectives become removed after an infectious period which is distributed as a random variable with an Exponential distribution with mean $1/\gamma$. Furthermore, a removed individual plays no further part in the epidemic and the latter ends when there are no more infectives in the population. All of the infectious periods and the infection process are assumed to be independent of each other.

\subsection{Bayesian Inference for Partially Observed Data}
Denote the set of removal times $0=R_1\leq R_2 \leq\cdots\leq R_n$; we assume that no removal times are unobserved and that the epidemic has run its course, that is the epidemic ends at time $R_n$. Denote by $I_i$ the infection time of the individual removed at time $R_i$, and let $I_{(i)}$ denote the $i$th ordered infection time. Let $\omega$ be the label of the first infective. Let $\mathcal{I}=\{I_1,\ldots,I_n\}$ and $\mathcal{R}=\{R_1,\ldots,R_n\}$; the joint density of $(\mathcal{I}_{-\omega},\mathcal{R})$ given $\beta$, $\gamma$ and $I_{\omega}$ is proportional to:
\begin{equation*}
\gamma^n\prod_{i=2}^n \beta X_{I(i)-} Y_{I(i)-} \exp\left\{-\gamma\sum_{i=1}^n(R_i-I_i)-\int_{I_\omega}^{R_n}\beta X_tY_tdt\right\} \chi(\mathcal{I},\mathcal{R}),
\end{equation*}
where $t-$ denotes the left limit and
\begin{equation*}
\chi(\mathcal{I},\mathcal{R}) = \left\{\begin{array}{ll}
						1 & \mbox{if $I_{(i+1)}\leq R_{i}$ $\forall i\in\{1,\ldots,n-1\}$}\\
						0 & \mbox{otherwise.}
						\end{array} \right.
\end{equation*}
The purpose of the function $\chi(\cdot, \cdot)$ here is to ensure that $\mathcal{I}$ and $\mathcal{R}$ are such that there is always at least one infective throughout the duration of the epidemic.

If $\mathcal{I}$ and $\mathcal{R}$ are known then inference for the infection rate $\beta$ and removal rate $\gamma$ is straightforward. However, temporal disease outbreak data typically consist only of the set of removal times $\mathcal{R}$. In this case the desired likelihood can be thought of as the integral, over all possible configurations of infection times, of the joint density $\pi(\mathcal{I}_{-\omega},\mathcal{R}\mid \beta,\gamma, I_{\omega})$. However, this integral is typically analytically and numerically intractable due to its high dimensionality and the non-trivial nature of the region of integration.

One way to overcome this is to use data-augmentation and introduce additional parameters which represent missing data in such a way that the likelihood becomes intractable. A natural choice of  augmentation is to use the unobserved infection times. Bayesian inference can then be drawn for both the unobserved infection times $\mathcal{I}$ as well as $\beta$ and $\gamma$ by employing a Markov Chain Monte Carlo algorithm to sample from the desired posterior distribution $\pi(\beta, \gamma, \mathcal{I} \mid \mathcal{R})$; see for example \citet{GibRen98} and \citet{OnRob99}.

\section{Bayesian Non-Parametric Estimation of the Force of Infection}
\subsection{Preliminaries}
We relax the parametric mass-action assumption in which the force of infection is proportional to the product $X_t Y_t$. We assume instead that it is an arbitrary function $h(t) > 0$ $(t \in \mathbb{R})$ that only depends on time and we wish to estimate within a Bayesian framework.

Conditional on knowledge of $h(\cdot)$, $\gamma$ and $I_\omega$, the augmented joint density of $(\mathcal{I}_{-\omega},\mathcal{R})$ is proportional to

\begin{equation}
\gamma^n\prod_{i=2}^n h\left(I_{(i)^{-}}\right) \exp\left\{-\gamma\sum_{i=1}^n(R_i-I_i)-\int_{I_\omega}^{R_n}h(t)dt\right\}\chi(\mathcal{I},\mathcal{R}). \label{eq:likelihood}
\end{equation}
We assume that $\gamma$ has a $\mbox{Gamma}(\kappa_\gamma, \mu_\gamma)$ prior distribution, i.e. it has mean $\kappa_\gamma/\mu_\gamma$ and that a priori $\omega$ is uniformly distributed on $1,\ldots,n$. Furthermore we assign to $R_1-I_{\omega}=-I_{\omega}$, which is the length of time between the first infection and first removal, an Exponential prior distribution with mean $1/\theta$. The remaining challenge is formulate a prior distribution for the function $h(\cdot)$ in such a way that that the terms in \eqref{eq:likelihood} - in particular its integral over $[I_{\omega},R_n]$ and the left limits at the infection times - are easy to compute.

\subsection{Prior probability models for $h(t)$: step function}
We first assume that the rate function $h(\cdot)$ on $[I_{\omega},R_n]$ is a step function. We now discuss how to formulate a prior distribution for it. Suppose there are $k$ changepoints $I_\omega<s_1<s_2<\cdots<s_k<R_n$ and the height on subinterval $[s_j,s_{j+1})$ ($j=0,\ldots,k$), where $s_0=I_{\omega}$ and $s_{k+1}=R_n$, is $h_j$ then we assume
\begin{equation*}
h(t)=\sum_{j=0}^kh_j1_{[s_j,s_{j+1})}(t).
\end{equation*}

With this form of $h(t)$, the total infectious pressure exerted from the infectives to the susceptibles over the course of the epidemic is trivial to compute:
\begin{equation*}
\int_{I_\omega}^{R_n}h(t)dt = \sum_{j=0}^kh_j(s_{j+1}-s_j).
\end{equation*}

We assume that a priori $k$ has a Poisson distribution with rate $\lambda$ conditioned on $k\leq k_{max}$, that $s_1,\ldots,s_k$ are distributed as the even-numbered order statistics of $2k+1$ points uniformly and independently distributed on $[I_{\omega},R_n]$. This prior choice for the changepoints of a step function was used by \citet{Green:1995} and serves the purpose of probabilistically spacing the changepoints by penalizing short intervals. The heights $h_0,h_1,\ldots,h_k$ have independent $\Gamma(\kappa,\mu)$ distributions, and $\mu$ itself a $\Gamma(a,b)$ prior.

As an alternative to assuming that the heights are a priori independent we can assume that they a priori follow a martingale structure. Following \citet{Arjas:1994}, we assume that $h_0\sim\mbox{Gamma}(\alpha_0,\beta_0)$ and that, given $h_0,\ldots,h_{i-1}$, $h_i\sim\mbox{Gamma}(\alpha_i,\beta_i)$ where $\alpha_i=\alpha$ and $\beta_i=\alpha/h_{i-1}$ so that $\mbox{E}[h_i\mid h_{i-1}]=h_{i-1}$. This should have the effect of smoothing $h(\cdot)$ in its average behaviour.

\subsection{Prior probability models for $h(t)$: 2nd-order B-spline}
We also consider as a prior model for $h(t)$ a 2nd-order B-spline, which is a continuous, piecewise quadratic.
We assume given $k+6$ knots $I_\omega=t_0=t_1=t_2<t_3<\cdots<t_{k+2}<t_{k+3}=t_{k+4}=t_{k+5}=R_n$, where the $k$ knots $t_3,t_4,\ldots,t_{k+2}$ are the interior knots, that $h(t)$ is a linear combination of B-spline basis functions
\begin{equation*}
h(t) = \sum_{i=0}^{k+2}P_{i+1}b_{i,2}(t),
\end{equation*}
where $b_{i,j}(t)$ is the $i$th B-spline basis function of order $j$. These basis functions can be defined recursively by
\begin{equation*}
b_{i,0}(t)=1_{[t_i,t_{i+1})}(t),
\end{equation*}
and
\begin{equation*}
b_{i,j}(t)=\frac{t-t_i}{t_{i+j}-t_i}b_{i,j-1}(t)+\frac{t_{i+j+1}-t}{t_{i+j+1}-t_{i+1}}b_{i+1,j-1}(t).
\end{equation*}
Thus,
\begin{align*}
h(t)=\sum_{j=2}^{k+2}1_{[t_j,t_{j+1})}(t)&\left[\frac{P_{j+1}(t-t_j)^2+P_{j}(t_{j+2}-t)(t-t_j)}{(t_{j+2}-t_j)(t_{j+1}-t_j)}\right.\\
&\left.+\frac{P_{j}(t_{j+1}-t)(t-t_{j-1})+P_{j-1}(t_{j+1}-t)^2}{(t_{j+1}-t_{j-1})(t_{j+1}-t_j)}\right].
\end{align*}
Furthermore,
\begin{equation*}
\int_{I_\omega}^{R_n}h(t)dt = \frac{1}{3}\sum_{j=1}^{k+3}P_j(t_{j+2}-t_{j-1}).
\end{equation*}

We assume that $k$ has an a priori Poisson distribution with rate $\lambda$ and maximum value $k_{max}$ and that the $k$ interior knots are distributed as the even-numbered order statistics of $2k+1$ points uniformly and independently distributed on $[I_\omega,R_n]$.

We need to ensure that the B-spline is non-negative. Since the basis functions are non-negative, a sufficient condition for this is that the coefficients $P_1,\ldots,P_{k+3}$ are non-negative. We fix $P_1=P_{k+3}=0$, so that the infection rate is $0$ at the start and end of the epidemic, and we assume that $P_2,\ldots,P_{k+2}$ have independent $\Gamma(\kappa,\mu)$ distributions, and give $\mu$ itself a $\Gamma(a,b)$ prior. B-splines with non-negative coefficients do not actually cover the full space of non-negative B-splines (which may have some negative coefficients), but they seem to cover the space fairly well and are much easier for computational purposes.

\subsection{Posterior computation}
Having assigned prior distributions to the quantities of interest, sampling from the posterior distribution of the parameters governing $h(\cdot)$, $\gamma$ and the unobserved infection times can be done via a Markov Chain Monte Carlo Algorithm. \vspace{0.4cm}

 {\em Step 1.} Sample $\gamma$ from conditional distribution $\Gamma\left(\kappa_\gamma+n,\mu_\gamma+\sum_{i=1}^n(R_i-I_i)\right)$; \vspace{0.4cm}

 {\em Step 2.} Update the parameters governing $h(t)$ using a Reversible-Jump Metropolis-Hastings algorithm; \vspace{0.4cm}

 {\em Step 3.} Update $\mathcal{I}$ using a Metropolis-Hastings algorithm.

We now describe in detail Step 2 for the case where the prior model specification for $h(t)$ is a step function assuming that heights $h_0, h_1, \ldots, h_k$ are independent. At each iteration of the algorithm we make one of three types of updates: 1. birth of a changepoint, 2. death of a changepoint, 3. within-model updates, i.e.~move existing changepoints and change heights. These occur respectively with probabilities $b_k$, $d_k$ and $1-b_k-d_k$, where
\begin{equation*}
b_k = c\min\{1,p(k+1)/p(k)\}, \quad d_{k+1} = c\min\{1,p(k)/p(k+1)\}
\end{equation*}
where $p(k)$ is the probability mass function of a Poisson random variable with rate $\lambda$ and maximum value $k_{max}$ and $c$ is chosen such that $\max_k(b_k+d_k)=0\cdot9$.

{\em Birth of a changepoint:} a new changepoint $s^*$ is proposed uniformly on $[I_\omega,R_n]$, which lies with probability $1$ in an existing interval, $(s_j,s_{j+1})$. We relabel $s_{j+1},\ldots,s_k$ as $s_{j+2},\ldots,s_{k+1}$ with heights relabelled accordingly and propose new heights $h'_j$ and $h'_{j+1}$ such that
\begin{equation*}
(s^*-s_j)\log(h'_j)+(s_{j+1}-s^*)\log(h'_{j+1})=(s_{j+1}-s_j)\log(h_j)
\end{equation*}
and
\begin{equation*}
\frac{h'_{j+1}}{h'_j}=\frac{1-u}{u}
\end{equation*}
where $u\sim\mbox{U}(0,1)$. The acceptance probability takes the form
\begin{equation*}
\min\{1,(\mbox{likelihood ratio})\times(\mbox{prior ratio})\times(\mbox{proposal ratio})\times(\mbox{Jacobian})\},
\end{equation*}
where the prior ratio is
\begin{equation}
\frac{p(k+1)}{p(k)}\frac{2(k+1)(2k+3)}{(R_n-I_\omega)^2}\frac{(s^*-s_j)(s_{j+1}-s^*)}{s_{j+1}-s_j}\frac{\mu^\kappa}{\Gamma(\kappa)}\left(\frac{h'_jh'_{j+1}}{h_j}\right)^{\kappa-1}\exp\{-\mu(h'_j+h'_{j+1}-h_j)\}
\label{eqn:birthpriorratio}
\end{equation}
and the proposal ratio is
\begin{equation*}
\frac{d_{k+1}(R_n-I_\omega)}{b_k(k+1)}
\end{equation*}
and the Jacobian is
\begin{equation*}
\frac{(h'_j+h'_{j+1})^2}{h_j}.
\end{equation*}

{\em Death of a changepoint:} a changepoint $s_{j+1}$ is selected uniformly at random from the existing changepoints for deletion and the new height over the interval $(s'_j,s'_{j+1})=(s_j,s'_{j+2})$ is determined by
\begin{equation*}
(s_{j+1}-s_j)\log(h_j)+(s_{j+2}-s_{j+1})\log(h_{j+1})=(s'_{j+1}-s'_j)\log(h'_j)
\end{equation*}
and the acceptance probability is obtained by inversion of the terms for the corresponding birth probability.

When we are not proposing a birth or a death, the following within-model updates take place, both of which can be repeated to improve mixing.

{\em Move an existing changepoint:} Provided $k>0$, a changepoint $s_j$ is selected uniformly at random from the existing changepoints and a new position $s'_j$ is proposed uniformly at random on $[s_{j-1},s_{j+1}]$ and accepted with probability
\begin{equation*}
\min\{1,(\mbox{likelihood ratio})\times\frac{(s_{j+1}-s'_j)(s'_j-s_{j-1})}{(s_{j+1}-s_j)(s_j-s_{j-1})}.
\end{equation*}

{\em Change a height:} a height $h_j$ is chosen uniformly at random from the existing heights and a new height $h'_j$ is proposed such that $\log(h'_j/h_j)$ is uniformly distributed on $[-1/2,1/2]$, and accepted with probability
\begin{equation*}
\min\{1,(\mbox{likelihood ratio})\times(h'_j/h_j)^\kappa\exp\{-\mu(h'_j-h_j)\}\}.
\end{equation*}

If the heights are assumed to be a-priori dependent according to the martingale structure described in the previous Section, then we can use the same mechanisms for updating the parameters, with changes to the acceptance probabilities for the birth or death of a changepoint and changing a height.

The acceptance probability for the birth of a changepoint changes only in the prior ratio, where $\mu^\kappa/\Gamma(\kappa)(h'_jh'_{j+1}/h_j)^{\kappa-1}\exp\{-\mu(h'_j+h'_{j+1}-h_j)\}$ in \eqref{eqn:birthpriorratio} is replaced by, if $j<k$,
\begin{equation*}
\frac{\alpha^\alpha}{\Gamma(\alpha)}\left(\frac{h'_j}{h_j}\right)^{\alpha_j-1-\alpha}\frac{1}{h'_{j+1}}\exp\left\{-\beta_j(h'_j-h_j)-\alpha\left(\frac{h'_{j+1}}{h'_j}+\frac{h_{j+1}}{h'_{j+1}}-\frac{h_{j+1}}{h_j}\right)\right\},
\end{equation*}
while if $j=k$,
\begin{equation*}
\frac{\alpha^\alpha}{\Gamma(\alpha)}\frac{\left(h'_k\right)^{\alpha_k-1-\alpha}\left(h'_{k+1}\right)^{\alpha-1}}{h_k^{\alpha_k-1}}\exp\left\{-\beta_k(h'_k-h_k)-\alpha\frac{h'_{k+1}}{h'_k}\right\},
\end{equation*}
with the death of a changepoint probability changing accordingly.

The acceptance probability for a change in height becomes, if $j<k$,
\begin{equation*}
\min\{1,(\mbox{likelihood ratio})\times(h'_i/h_i)^{\alpha_i-\alpha}\exp\{-\beta_i(h'_i-h_i)-\alpha h_{i+1}(1/h'_i-1/h_i)\}\},
\end{equation*}
while if $j=k$
\begin{equation*}
\min\{1,(\mbox{likelihood ratio})\times(h'_k/h_k)^{\alpha_k}\exp\{-\beta_k(h'_k-h_k)\}\}.
\end{equation*}

Step 3 involves updating the set of infection times inluding the label and time of initial infection.

{\em Update infection times:} an individual $j$ is selected uniformly at random from $\{1,\ldots,n\}\backslash{\omega}$ and a new infection time $I'_j$ is proposed such that $R_j-I'_j$ has an Exponential distribution with rate $\gamma$ and maximum value $R_j-I_\omega$. The proposed infection time is accepted with probability
\begin{equation*}
\min\{1,h(I'_j)/h(I_j)\}.
\end{equation*}

{\em Update first infection time:} a new first infection time $I'_\omega$ is proposed such that $\min(I_{(2)},s_1)-I'_\omega\sim\mbox{Exp}(\theta+\gamma+h_0)$. This proposal for $I'_\omega$ is restricted to values before the current second infection time and the first changepoint, the latter restriction ensuring the dimensionality is fixed within this update. The new first infection time is accepted with probability
\begin{equation*}
\min\left\{1,\frac{s_1-I'_\omega}{s_1-I_\omega}\left(\frac{R_n-I_\omega}{R_n-I'_\omega}\right)^{2k+1}\right\}.
\end{equation*}

{\em Update $\omega$:} the label of the first infective is updated by selecting a new value $\omega'$ uniformly at random from $\{i\in\{1,\ldots,n\}:I_i<R_\omega\}$, which is the set of individuals infected before individual $\omega$ was removed including $\omega$ themselves, and then setting new infection times $I'_\omega=I_{\omega'}$ and $I'_{\omega'}=I_\omega$.

Finally, the above changepoint algorithm can be used with minor adaptation when one assign a 2nd order B-spline prior for $h(\cdot)$.  Changepoints correspond to interior knots and heights to parameters and so all the updates to the B-spline are simply adapted from those updates to the step function with appropriate relabelling ($P_j$ for $h_j$, $t_j$ for $s_j$).

The only real adaptation we need make for updating infection times is for the update of the first infection time as follows: in this case we propose a new first infection time $I'_\omega$ such that $\min(I_{(2)},t_3)-I'_\omega\sim\mbox{Exp}(\theta+\gamma)$. The new first infection time is accepted with probability
\begin{equation*}
\min\left\{1,(\mbox{likelihood ratio})\times\exp\{-\gamma(I'_j-I_j)\}\frac{t_3-I'_\omega}{t_3-I_\omega}\left(\frac{R_n-I_\omega}{R_n-I'_\omega}\right)^{2k+1}\right\}.
\end{equation*}

\section{Results}
\label{sec:results}

\subsection{Application to simulated data}

We first illustrate the proposed methodology by applying it to synthetic datasets. Dataset 1 was generated using a mass action model for the infection rate, i.e.~$h(t)=\beta X_tY_t$, where $\beta=1\cdot7/1000$ and $\gamma=1$ with the epidemic started by one infective among a population of 1000 susceptibles. The number of individuals who were ever infective was $n=667$.

\begin{figure}
\includegraphics{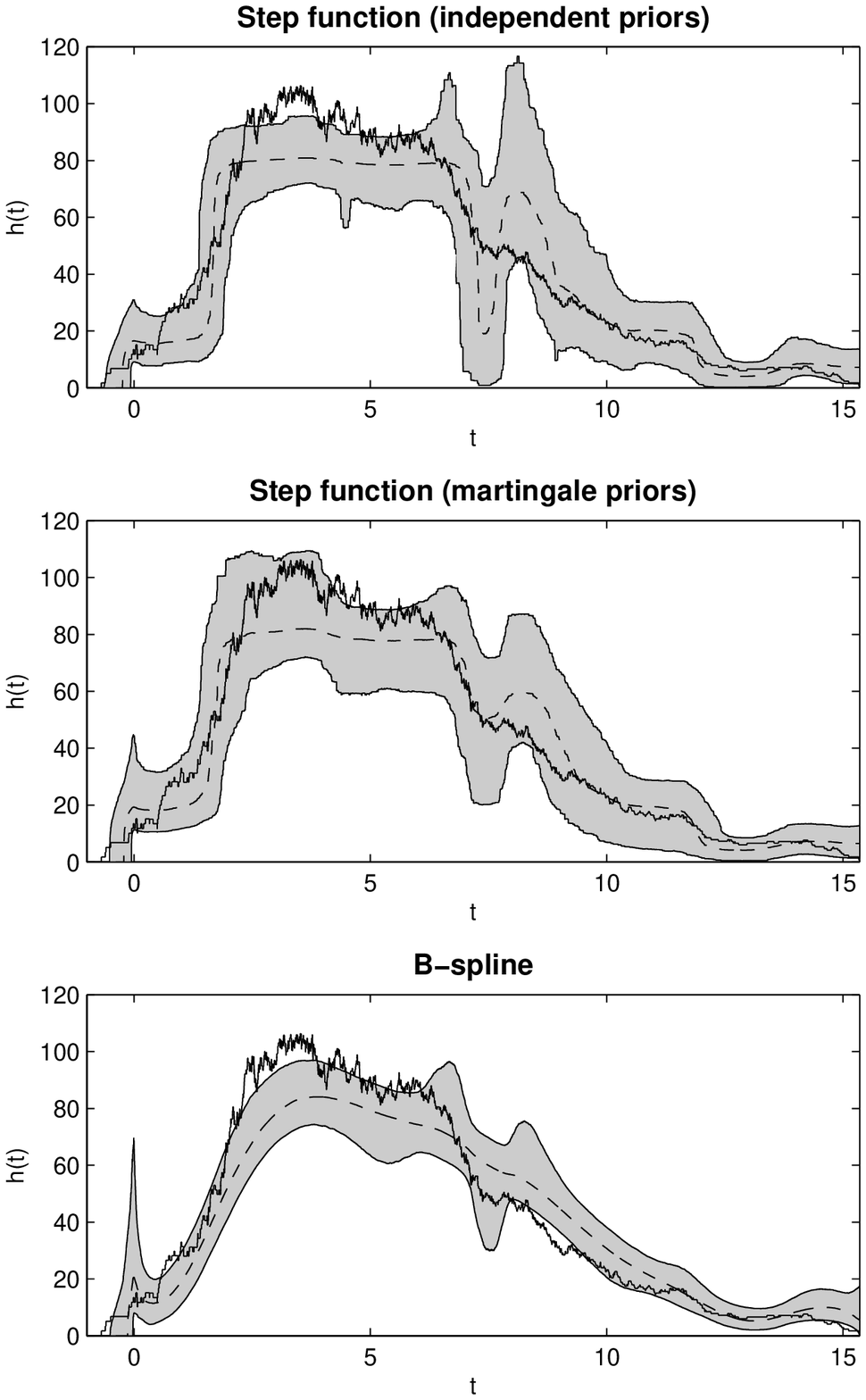}
\caption{Step function and B-spline models fitted to Dataset 1, with $\lambda=9$, $k_{max}=50$ and a vague prior distribution for $\gamma$. The dashed line is the median and the shaded area the 5th to 95th percentile of the posterior sample while the solid line is the true infection rate curve.}
\label{fig:XYdata_vaguegamma}
\end{figure}

\begin{figure}
\includegraphics{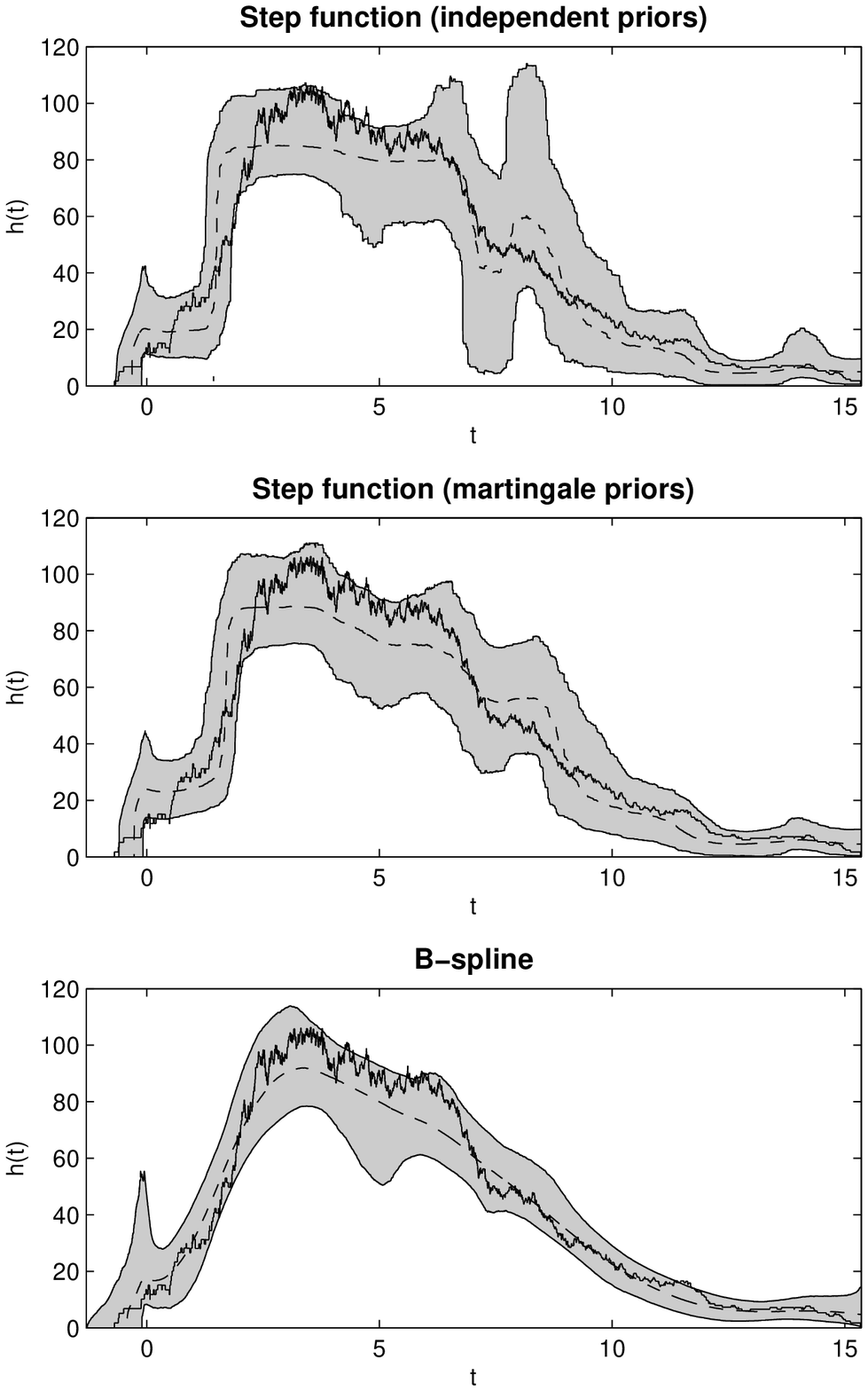}
\caption{Step function and B-spline models fitted to Dataset 1, with $\lambda=6$, $k_{max}=50$ and an informative prior distribution for $\gamma$. The dashed line is the median and the shaded area the 5th to 95th percentile of the posterior sample while the solid line is the true infection rate curve.}
\label{fig:XYdata_informativegamma}
\end{figure}

Figure \ref{fig:XYdata_vaguegamma} shows the results of fitting the non-parametric models to Dataset 1. We set $\lambda=10$ and $k_{max}=50$. All other parameters were set so as to be uninformative: $\theta=0$, $a=1$, $b=0$, $\kappa=1$, $\alpha_0=1$, $\alpha=1$, $\beta_0=0$ $\kappa_{\gamma}=1$ and $\mu_{\gamma}=0$. We see that the martingale priors induce smoothness in the average behaviour of the step function, leading to improved estimation. In all three cases there is clear underestimation at around $t=7\cdot5$ followed by overestimation to compensate at $t=8\cdot5$, which may be explained in part by the true infection rate levelling off slightly. However we noticed in all three cases that the sampled values of $\gamma$ tended to be much closer to $2$ than its true value of $1$, which leads to the sampled infection times generally being later than they should be. Figure \ref{fig:XYdata_informativegamma} then shows the results of instead using a more informative prior distribution for $\gamma$ by setting $\kappa_\gamma=n$ and $\mu_\gamma=n$. We see that this offers improved estimation, particularly for the B-spline.

Dataset 2 is generated from a modified mass action model with varying contact rate such that $h(t)=\beta(1+\cos(t-I_{(1)}))X_tY_t$, where $\beta=1\cdot7/10,000$ and $\gamma=1$ with the epidemic started by one infective among a population of 10,000 susceptibles. The number of individuals who were ever infective was $n=6971$.

\begin{figure}
\includegraphics[width=7in]{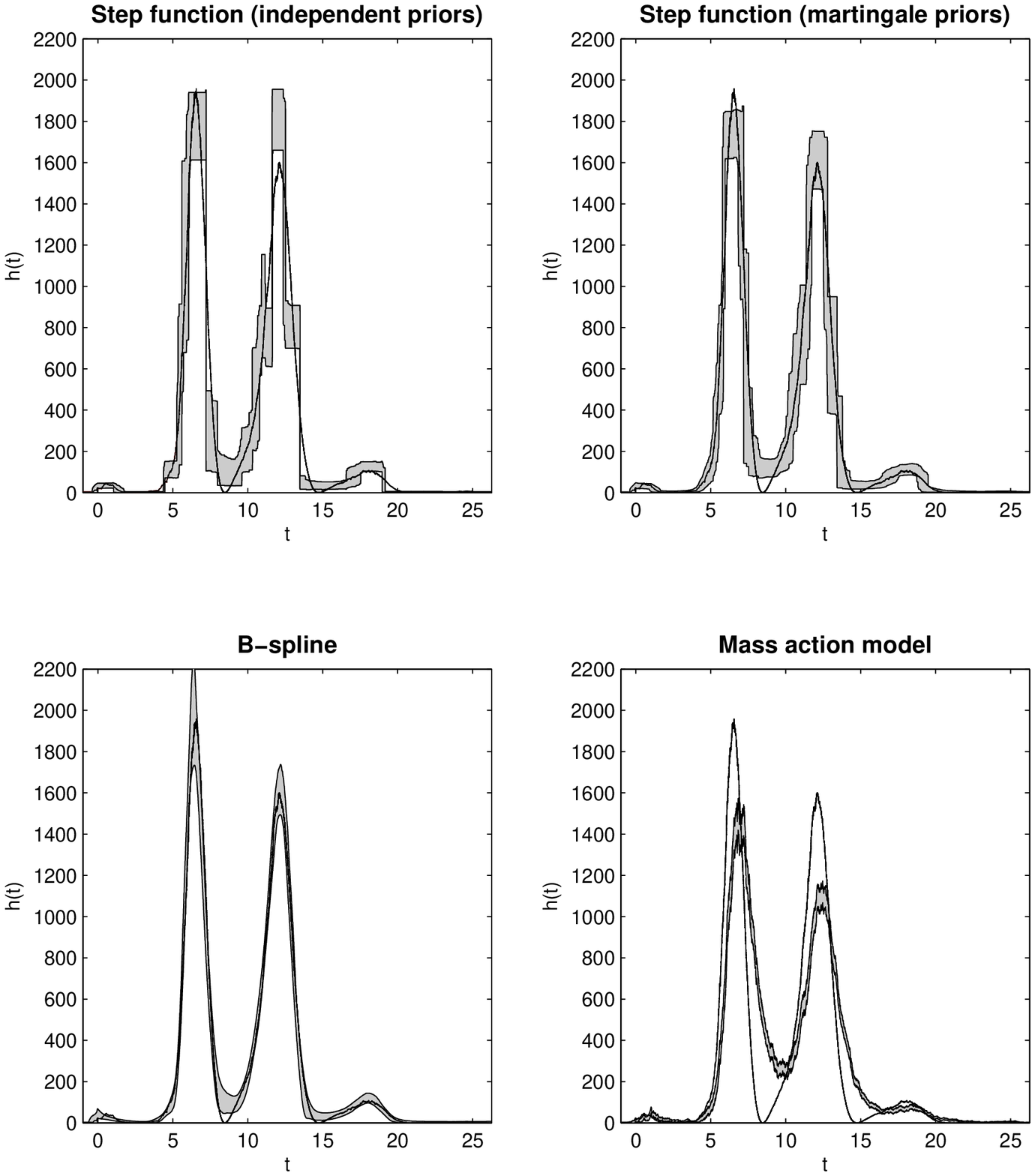}
\caption{Non-parametric models and the mass action model fitted to Dataset 2, with $\lambda=20$, $k_{max}=100$. The shaded area is the 5th to 95th percentile of the posterior sample while the solid line is the true infection rate curve.}
\label{fig:seasonal_10k_data_plots}
\end{figure}

Figure \ref{fig:seasonal_10k_data_plots} shows the results of fitting the non-parametric models to Dataset 2, where again we used the informative prior distribution for $\gamma$ by setting $\kappa_\gamma=n$ and $\mu_\gamma=n$. All other prior parameters were set so as to be noninformative as previously. For comparison we also fit the mass action model, although with a noninformative prior distribution for $\gamma$ in this case. The B-spline works very well and while the step functions are obviously less smooth, they manage to produce good results. With the mass action model, we see how fitting the incorrect model can lead to poor estimation of the infection rate.

\begin{table}[h]
\begin{center}
\caption{Typical runtime for the various models for these two datasets on a desktop.} \label{tab:runtime}
\begin{tabular}{lccc}
& {\em Step function} & {\em B-spline} & {\em Mass action model} \\
Dataset 1 ($n=657$) & 7 mins & 12 mins & 29 mins\\
Dataset 2 ($n=6971$) & 2 hours 10 mins & 3 hours 30 mins & 24 hours\\
\end{tabular}
\end{center}
\end{table}

Table \ref{tab:runtime} shows the typical runtimes for the different models when fitted to Datasets 1 and 2. From this we might expect that for a dataset of around 65,000-70,000 removal times, that the step function and B-spline would take about a day or so to run, while the mass action model could take several weeks. The main reason the computational advantage of the non-parametric models over the parametric ones is that infection time updates (beside that of the initial infective) do not also update the infection rate curve, which they do for the mass action model. The step function algorithms run more quickly than the B-spline one as calculation of the infection rate at the infection times is much simpler. The best model to fit then would seem to be determined by the trade-off between the speed of a step function with martingale priors and the better, smoother fit of the B-spline.

\subsection{Application to 2003 Hong Kong SARS symptom onset data}

\begin{figure}
\includegraphics{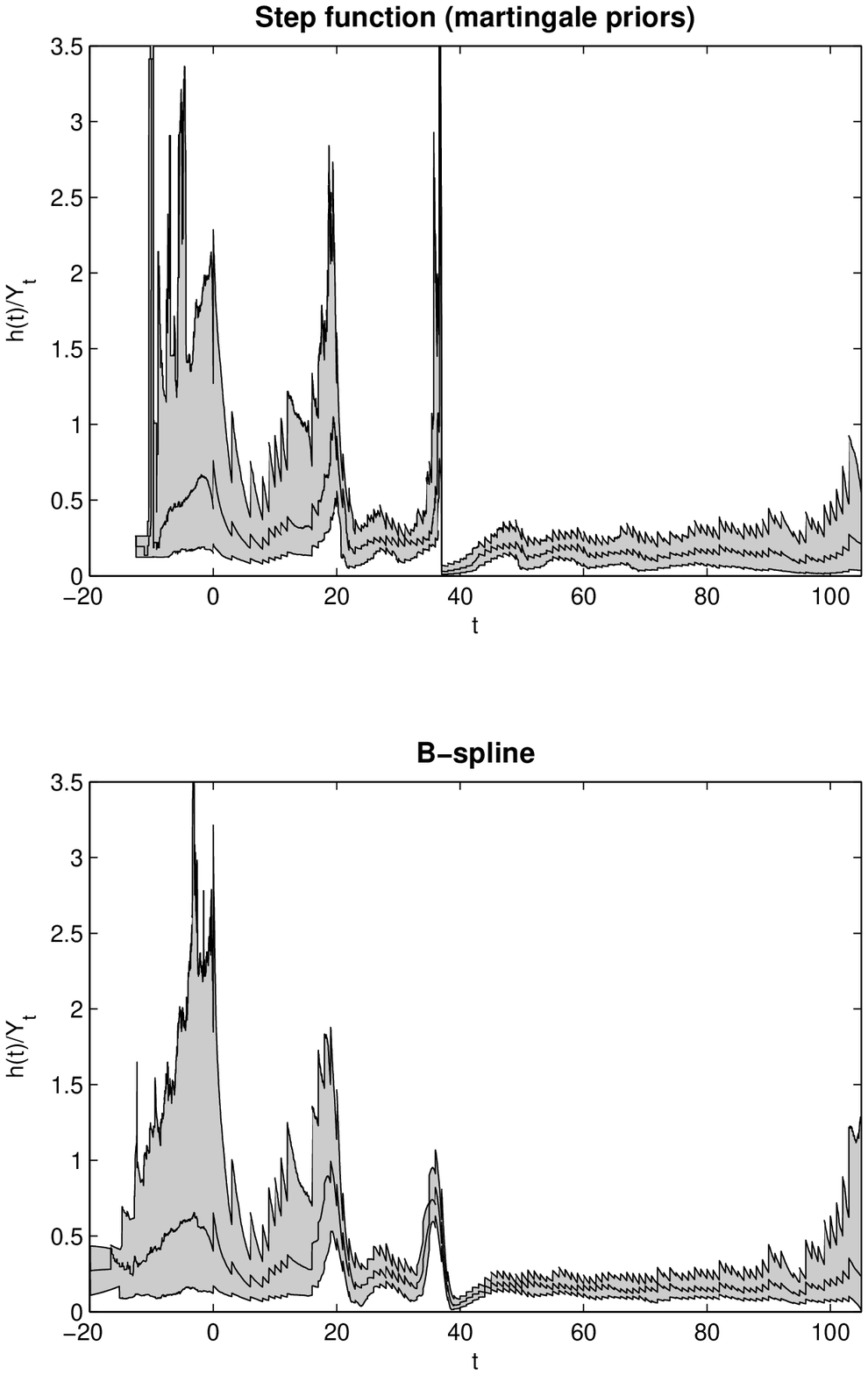}
\caption{Infection rate-per-infective for the Hong Kong SARS data (time scale is in days). The solid line is the median and the shaded area the 5th to 95th percentile of the posterior sample.}
\label{fig:sars_data_infection_rate_per_infective}
\end{figure}

We now fit the non-parametric models to data from the 2002-03 severe acute respiratory syndrome (SARS) outbreak. The data consist of the symptom onset dates of all 1755 patients identified in Hong Kong \citep{Leung:2004}. An advantage of our model is that we do not need to know or estimate the size of the population as there is no need to keep track of the number of susceptibles throughout the epidemic (the population size of Hong Kong is sufficiently large that the number of susceptibles remains of the same order throughout the epidemic). Results for fitting the step function with martingale priors and the B-spline are presented in Fig.~\ref{fig:sars_data_infection_rate_per_infective}, where we have set $\lambda=20$, $k_{max}=100$, $\kappa_\gamma=n$ and $\mu_\gamma=4\cdot6n$ because $4\cdot6$ days is the mean time from infection to onset of symptoms according to \citet{Leung:2004}. A symptom onset date is not truthfully equivalent to a removal time as a symptomatic individual can still spread infection e.g.~to hospital workers, however the results were fairly similar when we introduced fixed-length symptomatic infectious periods of 15, 20 or 25 days. A more rigorous analysis of this dataset may want to use time of death as a removal time for individuals who died and some given distribution for the symptomatic infectious period of infected individuals who survived. There appears to be a tendency for greater uncertainty at the beginning and end of the epidemic. There are two very identifiable spikes in the spread of the epidemic, one about 20 days after the first symptom onset, the other after about 37 days. These correspond to two previously identified "super-spread" events (SSEs) (\citep{Riley:2003}), the first in which a large number were infected by the index patient in the Prince of Wales Hospital, the second in which a cluster of infections occurred at the Amoy Gardens estate. It is impressive that the effect of these two events has been picked up by our models without explicitly incorporating them.

\section{Discussion}
\label{sec:discussion}

Non-parametric modelling could be used to help indicate if any particular parametric model fits well. For instance, if the curve of the average infection rate-per-infective/susceptible pairing is flat, this would indicate the mass action model fits well. Plotting average infection rate-per-infective should be sufficient if the population is large relative to the number of infections. If instead the curve rises to a peak in the middle before falling off to a similar level as at the start, then the infection rate may be proportional to the number of infectives squared, or some other power larger than 1. Non-parametric modelling could also be used to identify the effect of intervention methods. Not only could it indicate how much the intervention reduces the spread of the epidemic, but how quickly it does so too.

The methodology can be adapted to include other epidemic model assumptions. For instance, one may want to explicitly incorporate a latent period with some given distribution and place prior distributions on its parameters. However, a latent individual could be essentially considered an infective who does not contribute to the infection rate, and so, assuming the latent period is incorporated into the infectious period, our non-parametric models should still fit the data well. Also, while the model should be able to identify the overall infection rate for a household-based epidemic, one may want to, for instance, model the between-household infection rate non-parametrically alongside a fixed within-household contact rate.

The methodology can also be extended to semi-parametric models. For instance, one could assume that the infection rate at time $t$ takes the form $h(t)=\beta(t)X_tY_t$, and then model the contact rate $\beta(t)$ via a step function or B-spline. This would be essentially using the mass action model as a baseline, and in doing so could offer better estimation of the infection rate. The main drawback to this approach is that it would probably take at least as long to compute as the mass action model itself. Computation of the integral in \eqref{eq:likelihood} would also become more involved.

% references
\bibliographystyle{apalike}
\bibliography{refs}

\end{document}